\documentclass[aps,twocolumn,preprintnumbers,showpacs]{revtex4-2}

\usepackage{graphicx}  
\usepackage{dcolumn}   
\usepackage{bm}        
\usepackage{amssymb,amsmath,amsfonts}   
\usepackage{xcolor}
\usepackage{hyperref}
\hypersetup{
    colorlinks=true,
    linkcolor=blue,
    citecolor=blue,
    urlcolor=cyan,
    pdfpagemode=FullScreen,
    }

\def\rme{{\rm e}}
\def\rmd{{\rm d}}
\def\nn{\nonumber \\}
\def\gonetwo{\gamma_{12}}

\begin{document}

\title{Decay and revival dynamics of a quantum state embedded in 
regularly spaced band of states}

\author{Jan Petter Hansen}
\email{jan.hansen@uib.no}
\author{Konrad Tywoniuk}
\email{konrad.tywoniuk@uib.no}
\affiliation{Department of Physics and Technology, University of Bergen, 5020 Bergen, Norway}

\date{\today}


\begin{abstract}
The dynamics of a single quantum state embedded in one or several (quasi-)continua is one of the most studied phenomena in quantum mechanics. In this work we investigate its discrete analogue and consider short and long time dynamics based on numerical and analytical solutions of the Schr\"odinger equation. In addition to derivation of explicit conditions for initial exponential decay, it is shown that a recent model of this class  [Phys. Rev. A 95, 053821, (2017)], describing a qubit coupled to a phonon reservoir with energy dependent coupling parameters is identical to a qubit interacting with a finite number of parallel regularly spaced band of states via constant couplings. As a consequence, the characteristic near periodic initial state revivals can be viewed as a transition of probability between different continua via the reviving initial state. Furthermore, the observation of polynomial decay of the reviving peaks is present in any system with constant and sufficiently strong coupling. 
\end{abstract}

\maketitle

\section{Introduction}
\label{sec:intro}

The decay dynamics of an unstable or excited quantum state has been one of the most studied topics since
 the invention of quantum mechanics. Following the pioneering works of Dirac and Landau almost 100 years
 ago \cite{dirac1927quantum, landau1927dampfungsproblem}, Wigner and Weisskopf \cite{weisskopf1930calculation} derived the characteristic Lorentzian line-shape of the final state probabilities assuming exponential decay of the initial state. The constant transition rate derived in the earliest works was later known as Fermi's ``Golden Rule'' with reference to his ground-breaking description of nuclear $\beta$-decay \cite{wilson1968fermi}. In the period up till today the number of articles addressing the topic is probably in the order  of thousands or more, when counting decay phenomena in subatomic, atomic, molecular and solid state physics. 
 
In the time-dependent regime a constant transition rate, as provided by the ``Golden Rule,'' immediately leads to an exponential decay law using basic statistics of independent events together with a constant rate \cite{merzbacher1998quantum}. However, there exist numerous cases where non-exponential dynamics is observed, e.g., in \cite{rothe2006violation,peshkin2014non,kelkar2004hidden,andersson2019non} and indeed the systems where the Hamiltonian provides an exact exponential decay law directly from the time-dependent Schr\"odinger equation, are quite few.

The prime example of exact exponential decay from initial time $t=0$ up to a fixed time $t+\Delta t$ appears in a model system with constant coupling from the initial embedded state to a regularly spaced band of states, sometimes representing a discretized continuum. In its simplest form there is no coupling between band states \cite{milonni1983exponential,cohen1998atom}. Thus, the interaction matrix contains non-zero elements on a single row and column which motivates to refer to this model in following as the row-column (RC) model \cite{stey1972decay}. The RC model dates back to the the 1930s and was analyzed by Fano \cite{fano1935sullo} in 1935. Examples of real systems that can be reasonably well described by the RC model are excited atoms or molecules decaying through one-photon transitions to lower state or electron spins interacting with a bath of nuclear spins \cite{coish2008exponential}. Very recently, it was shown that initial exponential decay is a universal phenomenon occurring in a wide range of related many body systems \cite{micklitz2022emergence}. It was also shown \cite{guo2017giant} that harmonic energy dependent coupling parameters results in initial exponential decay of an artificial two-level atom, interacting with surface acoustic phonon states. 

\begin{figure*}[t] 
\begin{center}
\includegraphics[width=0.75\textwidth]{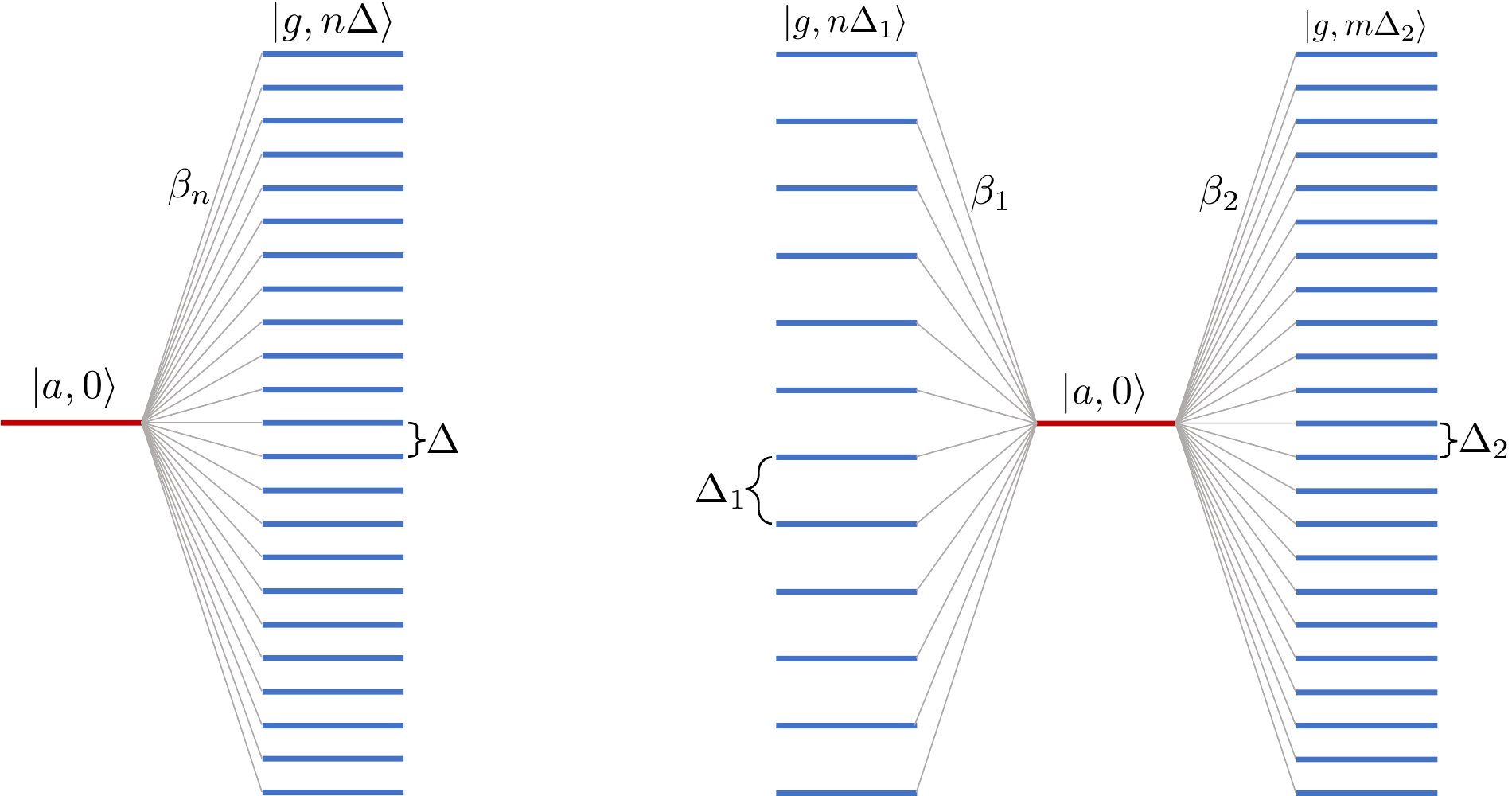}
\caption{Schematic presentation of the models studied in this work.  Left: The state $|a,0\rangle$ embedded in a infinite band of states $|g,n\Delta \rangle$ with band energy separation
$\Delta$ and coupling $\beta_n$. 
Right: The state $|a,0\rangle$ embedded in two (or more) bands of states characterised with different band energy separations and constant coupling to each
state of the band. 
 }
\label{fig0:models}
\end{center}
\end{figure*}

An RC model with regularly spaced energy band, with energy separation  $\Delta$, reproduce the dynamics of a true continuum as long as the product of the absolute square of the coupling parameter $|\beta|^2$ and density of states  $\Delta^{-1}$ are kept constant in the limit where $\Delta \rightarrow 0$ ($\Delta$ itself represents the energy separation between the states in the band, which we assume for now are uniformly distributed). The discrete and the continuum system then reproduce the same physics up to a maximum time $\tau = 2\pi/\Delta$. From then on the dynamics of the discrete system separate from its continuous counterpart. A series of revivals of the initial state probability occur in the discrete version at times near integer number of $\tau$. The revival phenomenon of the RC model was initially discussed in \cite{yeh1982interrupted} and is also well known in other discrete systems, such as the Jaynes-Cummings model in quantum optics \cite{eberly1980periodic}.

The origin of the revivals of the RC model can be understood from the fact that the dynamics of the initial state may be transformed into a time delay equation \cite{milonni1983exponential}. Each time the system passes through an integer number of $\tau$ a new term is added to the amplitude which can cause a partial revival. Alternatively the sequence of imperfect revivals can be understood from the quasi-periodicity which characterizes this particular model when an eigenvector basis is applied. In \cite{guo2017giant}  it was shown that revivals also occur with harmonic coupling to a true continuum, in sharp contrast to the infinite value of $\tau$ for constant coupling. In this case the revivals are occurring at the time-scale $T$ of the harmonic modulation of the coupling. The phenomenon was later confirmed in experiments \cite{andersson2019non}.

In this paper we explore the short and the long time dynamics in the RC model for both types of couplings, constant and harmonic, as well as coupling two distinct bands. The analysis and computations will be based on the eigenvector basis of the full Hamiltonian, which gives a completely analytical expression for the dynamics as long as the energies of the eigenstates are numerically obtained \cite{cohen1998atom,gaveau1995limited}. A coherent theoretical analysis, based on the Laplace transform method \cite{stey1972decay}, valid for both coupling types and independent of the inverse density of states is outlined in parallel. We will show that there exists a close connection between constant couplings to  parallel bands and the harmonic coupling to a single continuum. This observation leads to the identification of a long-time power law decay of revival peaks in the constant-coupling system which is identical to what was discovered in the harmonic case. By simply including an additional dense background reservoir to account for, e.g., noise and temperature fluctuations, we obtain excellent agreement with the experiment \cite{andersson2019non}. The analysis is carried out in the two following sections, one for each coupling type. The Laplace transform method is is provided an Appendices. Atomic units are applied throughout.

\section{The RC model with constant coupling}
\label{sec:rc-constant-coupling}

Most theoretical methods for solving the Schr\"odinger equation in a true continuum are based on various analytical methods to obtain a closed form expression of the amplitude describing the embedded decaying initial state. Two classic approaches are the ones of Stey and Gibberd \cite{stey1972decay}, utilizing the Laplace transform, and that of Milonni et al. \cite{milonni1983exponential}, based on the Poisson summation formula. In both cases a delayed time differential equation of the initial amplitude appears. Our approach is based on a discrete representation of the continuum, diagonalizing the Hamiltonian and expressing the initial state in the eigenvector basis. The true continuum dynamics can be obtained when letting the energy separation go to zero in the final expression $\Delta \to 0$ while keeping the ratio $|\beta|^2/\Delta =$ const, where $\beta$ is the coupling between the singled-out state $a(t)$ and the band. This approach has some advantages since the eigenvector basis is analytical and also numerically available on almost any computer for $10^4$ states or more. Computations of time development becomes extremely efficient since only eigenvectors which has a significant overlap with the initial state needs to be included in the basis. 

\subsection{A single state embedded in a single quasi-continuum}
\label{sec:single-continuum}

Consider an unstable state $| a, 0 \rangle$ with energy $\varepsilon_a$ which decay to the ground state  with energy $\varepsilon_b$ via interactions with a (quasi-)continuum. The field free transition frequency is defined as $\omega_{ab} = \varepsilon_a-\varepsilon_b$. Accompanying the initial state is a vacuum state of the continuum, indicated by the zero in the initial state ket-vector. Correspondingly, the transition to the ground state imply that some of the continuum states are populated by an energy quanta $ \varepsilon_n = n \Delta$. For a slow transition only $\varepsilon_n=\omega_{ab}$ gets populated after infinitely long times. For finite transition times many eigenvectors may overlap with the initial state. The state vector describing the complete single photon (phonon) transition, $|\Psi(t) \rangle$, is then described by the superposition, 
 \begin{equation}
| \Psi(t) \rangle = a(t) | a, 0 \rangle  + \sum_{n} b_n(t)  | g, n \Delta \rangle, 
\end{equation} 
with $a(t)$, $b_n(t)$ being time dependent probability amplitudes. By assuming that the state $|a\rangle $ couples with strength $\beta_i$ only to each of the discretized states and that the latter do not couple between themselves, the time dependent Schr\"odinger equation is transformed into a set of $N$ coupled first order differential equations for the expansion coefficients, 
\begin{equation}
\label{eq:H}
i \frac{\rmd}{\rmd t} \left(
\begin{array}{c}
a \\ {\bf b}
\end{array}\right)
=  
\left(
\begin{array}{cc}
\omega_{ab} & {\bf C}^{\dagger} \\ 
{\bf C} & {\bf \Omega} \\
\end{array}
\right)
\left(
\begin{array}{c}
a \\ {\bf b}
\end{array}\right) \,.
\end{equation}
Here the row vector  ${\bf b}$ contains the $N-1$ expansion coefficients $b_n(t)$. The sparse coupling Hamiltonian is Hermitian with nonzero elements defined by the row vector  ${\bf C}^{\dagger} = \{ \beta_1^*, \beta_2^* ... \beta_{N-1}^*\}$  and the energy levels of the discretized continuum are given by the diagonal matrix, ${\bf \Omega} = {\rm diag}\{\varepsilon_1,  \varepsilon_2... \}$. It is straightforward to solve Eq.~\eqref{eq:H} numerically even for order of thousands to hundreds of thousands of equations. Alternatively, and in general more computationally efficient, we may diagonalize the coupling matrix once, and obtain the time dependent amplitudes explicitly.  The survival amplitude takes the form, 
\begin{equation}
\label{eq:a(t)}
a(t) = \langle a | \Psi(t) \rangle    =    \sum_{n=-\infty}^\infty |\langle a | v_n  \rangle|^2  | v_n \rangle e^{-iE_n t} \,,
\end{equation}
where $| v_n \rangle$ is the $n$'th eigenvector having energy $E_n$ and $a(0)=1$. The simple structure of the coupling matrix leads to analytical eigenvectors and eigen-energies which can be obtained from a simple implicit equation for each eigenvalue, 
\begin{equation}
\label{eq:implicit}
E_m-\omega_{ab} = \sum_{n=-\infty}^\infty \frac{|\beta_n|^2}{E_m-\varepsilon_n} \,.
\end{equation}
As the right-hand side has poles at $E_m = \varepsilon_n$, we obtain exactly one eigenvalue in each interval 
$(\varepsilon_n,\varepsilon_{n+1})$. Furthermore, the set of energies $E_m-\omega_{ab}$ is symmetrically
distributed around $\omega_{ab}$.  In the case of an unbounded continuum the effect of the
detuning is simply to shift the distribution populated final amplitudes $b_n$ around centre eigenvalues from $E_m \sim \omega_{ab}$ to $E_m \sim 0$. 
Without loss of generality we therefor ignore the detuning in the following. 
The positive eigen-energies can then be expressed as $E_m = m + \delta_m$ for integer $m \ge 0 $ and  we have $E_{-m} = E_m$ \cite{cohen1998atom}. With the eigenvalues at hand, we obtain the eigenvectors
\begin{equation}
{\bf{v}}_m = \frac{1}{x_m}
\left(
\begin{array}{c}
1 \\  \frac{\beta_1}{E_m-\varepsilon_1} \\   \frac{\beta_2}{E_m-\varepsilon_2} \\ . \\  . \\  . 
\end{array}
\right) \,,
\end{equation}
with
\begin{equation}
\label{eq:x_i2}
 x_m^2 =  1 + \sum_{j=-\infty}^\infty \frac{|\beta_j|^2}{(E_m - \varepsilon_j)^2} \,.
\end{equation}
In \cite{cohen1998atom} it is shown that when all couplings elements are identical, $\beta_j = \beta$ for all $j$, 
the summation above can be made, yielding
\begin{equation}
\label{eq:sin}
 \sum_{j=-\infty}^\infty \frac{|\beta|^2}{(E_m - \varepsilon_j)^2} = 
 \left[\frac{\pi\beta}{\Delta\,  \sin\left(\pi \frac{E_m}{\Delta}\right)}\right]^2
\end{equation}
It follows that the implicit eigenvalue in Eq.~\eqref{eq:implicit} can be written as
\begin{equation}
E_m =  \frac{\pi\beta^2}{\Delta \tan \left(\pi \frac{E_m}{\Delta}\right)}  \,.
\end{equation}
The normalization factor $x_m$ can then be brought to the form,
\begin{equation}
\label{Tan1}
 x_m^2 =  \frac{1}{\beta^2} \left[ \beta^2 + \left(\frac{\pi \beta^2}{\Delta}\right)^2 + E_m^2 \right],
\end{equation}
which leads to a survival probability expressed as,
\begin{equation}
a(t) =  \sum_{n=1}^\infty \frac{2 \beta^2}{\beta^2 + (\gamma/2)^2 + E_n^2}\cos(E_n t) \,,
\label{eq:dyn_a0}
\end{equation}
where we introduced the rate $\gamma \equiv 2\pi\beta^2/\Delta$. 

Exponential decay is obtained in the limit of $\Delta \rightarrow 0$ while requiring that the rate $\gamma$ is kept constant. The first term in the denominator
then vanish and we can take the continuum limit of the sum over band energies. The resulting integral over energy gives exponential decay for all times,
\begin{equation}
a(t) = \frac{\gamma}{\pi} \int_0^{\infty} \rmd E \, \frac{\cos(E t)}{(\gamma/2 )^2 + E^2}    = 
\rme^{-t \gamma/2} \,.
\label{eq:dyn_a1}
\end{equation}
Hence, the probability of the state obeys Fermi's ``Golden Rule,'' namely $P_a(t) = \left| a(t) \right|^2 = \rme^{-t \gamma}$.

\begin{figure*}[t] 
\begin{center}
\includegraphics[width=0.75\textwidth]{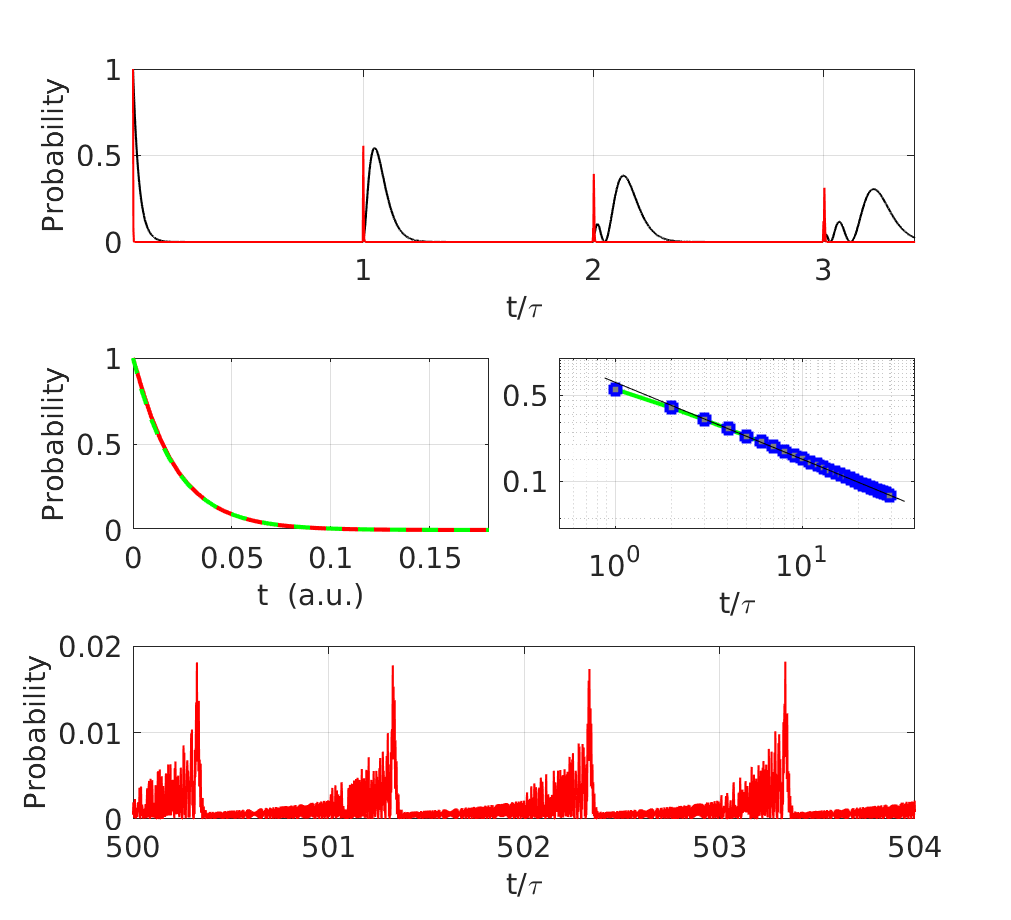}
\caption{The upper panel displays the probability for population of the initial state, $|a(t)|^2$ of Eq. (\ref{eq:dyn_a2}) for two sets of coupling parameters: $(\beta,\Delta)$ = (0.07,0.035) (black curve) and 
$(\beta,\Delta)$ = (0.2496,0.2048) (red curve) for a timespan covering the three first revivals. The middle panels shows the exponential decay of the red curve together with the analytical result of
 Eq. (\ref{eq:dyn_a1}) (green broken line) to the left.
The right middle panels displays the value of the first 30 revival peaks as blue squares connected by a green line. 
A straight line (black curve) is augmented for comparison with exact polynomial decay. 
The lower panel shows the time development of the red parameter set for a time region after 500 revivals. }
\label{fig:const-coupling}
\end{center}
\end{figure*}

For true discrete systems the replacement of the sum with an integral is less straightforward. The coupling $\beta$ may well be much larger than $\gamma$ in Eq.~\eqref{eq:dyn_a0} which prohibits the transition to Eq.~\eqref{eq:dyn_a1}. Thus, one may be lead to the conclusion that the decay dynamics is entirely non-exponential. However, there is always a finite time period $t \in (0, \tau)$, with $\tau \equiv 2\pi/\Delta$, where the decay follows Eq.~\eqref{eq:dyn_a1} exactly. Large $\Delta$, or a small rate $\gamma$, imply then that the decay is only partial since $\left|a(\tau) \right|^2 > 0$.   

The proof of the preceding statements is simple: A system with fixed and finite separation between the energy levels can, in the period $t \in (0,\tau)$, be replaced with a factor $k$ denser quasi-continuum. The decay dynamics of both systems are then exactly equal. In the dense band the coupling is given by $\beta_k = \beta / \sqrt{k}$. Since $k$ can be taken arbitrarily large the factor $\beta_k^2$ in the denominator eventually vanish and the validity of the transition from Eq.~\eqref{eq:dyn_a0} to \eqref{eq:dyn_a1} is restored up to $t = \tau$. For $t > \tau$ the dynamics of the original (with spacing $\Delta$) and dense (with spacing $\Delta_k$) systems diverge. 

The general amplitude for the embedded state of a true discrete system can now be written in terms 
of $t = m \tau + t_m$, 
\begin{equation}
a(t) =   \sum_{n} \frac{\beta^2}{\beta^2 + \left(\frac{\pi \beta^2}{\Delta}\right)^2 + 
E_n^2}e^{-i m \delta_n \tau} e^{-iE_n t_m} \,, 
\label{eq:dyn_a2}
\end{equation}
where $0<t_m <\tau$. 
This expression is identical to the solution based on delayed time differential equations below. 

In Fig.~\ref{fig:const-coupling} we display some characteristics of the solution. In the upper panel we observe the characteristics of initial dynamics for two different set of coupling parameters and densities of states, both plotted in scaled time units $t/\tau$. The revivals occur initially close to integer times, the height of the peak of the revival probability seems to be independent of the coupling strength while the ``lifetime'' of the revival region goes from wide to spike-like as the coupling strength is gradually increased. No matter how large the coupling becomes, the initial decay is always exponential and given by Eq.~\eqref{eq:dyn_a1}, as illustrated in the middle left panel of Fig.~\ref{fig:const-coupling}. Likewise, collecting the peaks in the strongest coupling case (red curves) and plotting them in a log-log plot we observe a near polynomial decay for a large number of peaks (from the second revival and outwards in time). Finally, to be discussed below, we plot the dynamics after a large number of revival times (500) in the lowermost panel. Here we see an almost chaotic dynamics building up towards the max revival peak. The position of the peak on the time axis has now been clearly shifted beyond the integer numbers (500, 501..) of $\tau$. At even larger numbers of $\tau$ eg. $10^4$, the signature of the revivals gets more and more blurred and will, due to the true non-periodicity of the system, never return towards their initial prominence. 

In App.~\ref{sec:laplace}, we also derive the time-evolution using directly the coupled set of evolution equations and solving them using the Laplace transform. In App.~\ref{sec:app-const-coupling}, we obtain the following solution for the time evolution of state $a(t)$,
\begin{equation}
\label{eq:const-sol-1}
a(t) = a_0(t) + \sum_{k=1}^\infty a_k(t) \,,
\end{equation}
where $a_0(t) = \rme^{- t \gamma/2}$, and
\begin{equation}
\label{eq:ak-final}
a_k(t) = - \frac{\gamma t_k}{k}\rme^{- t_k\gamma/2} L_{k-1}^{\{1\}}\big(\gamma t_k\big) \Theta\big(t_k \big) \,,
\end{equation}
where we defined $t_k \equiv t-k\tau$ and $L_{n}^{\{\alpha\}}(x)$ is the generalized Laguerre function, see also \cite{stey1972decay}. In particular, the first terms read
\begin{align}
\label{eq:a1-constant}
a_1(t) &= -\gamma \left(t-\tau\right)\rme^{-(t-\tau)\gamma/2} \Theta\big(t-\tau \big) \,,\\
\label{eq:a2-constant}
a_2(t) &= \left[ - \gamma (t-2\tau) + \frac{\gamma^2}2 (t-2\tau)^2\right] \nonumber\\
&\times\rme^{-(t-2 \tau)\gamma/2} \Theta\big(t- 2 \tau \big) \,,\\
\label{eq:a3-constant}
a_3(t) &= \left[ - \gamma (t-3\tau) + \gamma^2 (t-3\tau)^2 - \frac{\gamma^3}{6}(t-3\tau)^3 \right] \nonumber\\ &\times \rme^{-(t-3 \tau)\gamma/2} \Theta\big(t-3\tau \big) \,.
\end{align}
At first look, it seems sufficient to demand that
\begin{equation}
\tau \gamma/2 \gg 1 \,,
\end{equation}
in order to ensure that each term decays sufficiently fast, and in the step $n \tau < t < (n+1) \tau$ it is only the $n$th term that contributes. We will assume this condition is fulfilled in the following discussion.

The first contribution $a_1(t)$, that breaks with the exponential decay behavior, sets in at times $ \tau < t < 2\tau$, and reaches a peak value at $t = \tau + 2/\gamma$. Here, the probability peaks at $4/\rme^2\approx 0.54$. In the second interval, $2\tau < t < 3\tau$, $a_2(t)$ starts contributing leading to two peaks at times $t = 2\tau +\big( 3 \pm \sqrt{5}\big)/\gamma$. At these positions, the probability peaks at $\{4(2-\sqrt5)^2\rme^{-3+\sqrt5}, 4(2+\sqrt{5})^2 \rme^{-3-\sqrt5} \} \approx \{0.10, 0.38 \}$, respectively. Hence, the smaller peak precedes the bigger one. Similarly, in the third interval, $3\tau < t < 4 \tau$, there are three peaks and so on.

There are several features worth pointing out. First, while the position of the peaks clearly depend on $\tau$ and $\gamma$, the height of the peaks are completely independent of the parameters of the model. Second, at late times the multiple peaks that occur force the largest, and also latest, peak to be displaced so that ultimately the condition $\tau \gamma/2 \gg 1$ is not sufficient to ensure that only one contribution from the sum contributes in a certain time interval. This ``leakage'' effect inevitable leads to a chaotic behavior at late times, as can be inferred from the bottom panel of Fig.~\ref{fig:const-coupling}.

\subsection{Two and more parallel bands of states} 
\label{sec:const-two-bands}

In nature, several de-excitation processes occur in the presence of competing and separable continua 
such as Auger versus radiative decay processes in atomic physics. The problem was treated by Fano in 1961 \cite{fano1961effects} for real continua. Here we consider the dynamics for parallel energy bands with focus on the total system behavior when the involved continua have different isolated revival times.  Within the RC model, two-continua in its simples form, has two different constant coupling parameters, $\beta_1$ and $\beta_2$ and/or inverse densities of states $\Delta_1$ and $\Delta_2$. No direct coupling between the bands are taken into account, as well as coupling between states belonging to the same band due via interaction with the initial state. The dynamics is now described by the expansion, 
\begin{equation}
| \Psi(t) \rangle = a(t) | a \rangle + \sum_{n} b_n(t)  | g, n\Delta_1 \rangle + \sum_{m}c_n(t)  | g, m\Delta_2\rangle \,,
\end{equation}
Without loss of generality, again we neglect the effect of detuning. Projecting out on each basis vector we obtain a coupling matrix which has the same structure as in the case of a single continuum, 
\begin{equation} 
i \frac{\rmd}{\rmd t} \left(
\begin{array}{c}
a \\ {\bf b} \\ {\bf c}
\end{array}\right)
=  
\left(
\begin{array}{ccc}
\omega_{ab} & {\bf C}_{1}^{\dagger} & {\bf C}_{2}^{\dagger} \\
{\bf C}_1 & {\bf \Omega_1} & {\bf 0} \\
{\bf C}_2 & {\bf 0}^{\dagger}  & {\bf \Omega_2}   \\
\end{array}
\right)
\left(
\begin{array}{c}
a \\ {\bf b} \\ {\bf c}
\end{array}\right) \,,
\label{eq:H2}
\end{equation}
where again the amplitudes and coupling to the bands has been expressed as vectors ${\bf b,c}$. The matrices ${\bf \Omega_{1,2}} $ contains the discrete band energy levels on the diagonal, ${\bf \Omega}_{1,2}^{n,m} = \delta_{n,m} n\cdot \Delta_{1,2}$. Finally the matrix ${\bf 0}$, contains only zeros since the two continua do not interact directly.

The implicit equation for the eigenvalues now becomes,
\begin{equation}
E_m = |\beta_1|^2 \sum_{n}\frac{1}{E_m-n\Delta_1} + |\beta_2|^2\sum_{k} \frac{1}{E_m-k\Delta_2 } \,.
\end{equation}
To obtain the decay dynamics, consider the case when $\Delta_2 > \Delta_1$ and replace this continuum with one which has the same energy separation as the first, i.e.  $\Delta_1$.  
This implies a reduced coupling strength in the replacement band $\hat{\beta}_2 = \beta_2 \sqrt{\Delta_1/\Delta_2}$. The new implicit equation for eigenvalues then takes the form, 
\begin{equation}
E_m = \left(|\beta_1|^2+|\hat{\beta}_2|^2\right) \sum_{n=-\infty}^\infty \frac{1}{E_m-n\Delta_1} \,.
\end{equation}
As remarked by Fano \cite{fano1961effects}, this leaves us with exactly the same identities Eqs.~\eqref{eq:implicit}-\eqref{eq:dyn_a1}. Therefore, the obtained exponential decay of the initial channel becomes,
\begin{equation}
P_a(t) = \rme^{- \gamma_{12}t} \,,
\end{equation}
at early times, where the rate constant 
\begin{equation}
 \frac{\gamma_{12}}{2} =  \pi \frac{|\beta_1|^2+|\hat{\beta}_2|^2}{\Delta_1}  
             =  \pi \left( \frac{|\beta_1|^2}{\Delta_1}+\frac{|\beta_2|^2}{\Delta_2} \right) \,.
\end{equation}
Each band become initially populated according to a weighted fraction
\begin{equation}
\label{eq:band-population}
P_{i}(t) = \frac{\gamma_i}{\gamma_{12}}\left(1 - \rme^{-\gamma_{12} t}\right) 
\end{equation}
where  $P_{i}(t)$ ($i=1,2$) is the total probability summed over all, in principle infinite, number of states 
within band $i$ and $\gamma_i =2\pi |\beta_i|^2/\Delta_i$, for further details see App.~\ref{sec:two-continua}.

\begin{figure*}[t] 
\begin{center}
\includegraphics[width=0.75\textwidth]{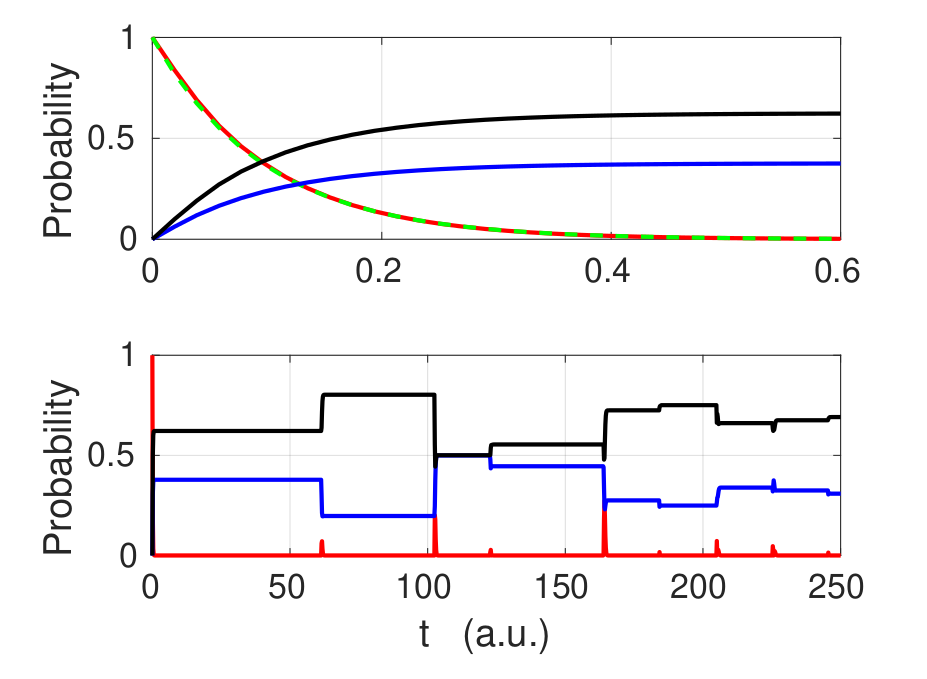}
\caption{Exponential decay and total population probability for two continua coupled
to the inital state with the same coupling parameter $\beta=0.2496$. The continuum represented with
the blue line has an energy separation $\Delta_1= 0.1024 \ (\tau_1=61.36)$ while the other has 
$\Delta_1= 0.0613 \ (\tau_2= 102.5)$. The upper panel shows the short time dynamics, the lower the long time dynamics.}
\label{fig:two-bands}
\end{center}
\end{figure*}

Exponential decay is now limited to the period $ t \in (0,\min[\tau_1, \tau_2])$, where $\tau_i = 2\pi/\Delta_i$. For integer revival periods of $\tau_1, \tau_2$ we will expect the system to undergo a back coupling of probability to the initial state as long as the band in question is not empty. In App.~\ref{sec:two-continua} we have solved the analytical case of two continua. In the strong coupling region a infinite set of additional revival times occur, namely at integer intervals of $\tau_1 + \tau_2$, as well as arbitrary multiples of the two, i.e. $n \tau_1 + m \tau_2$, where $n,m >1$ and $n\neq m$. 

Another striking feature, is that the heights of the revival peaks now depend on the ratio of the rates $\gamma_1/\gamma_2$. This,  in contrast to the case of constant coupling, and harmonic as we will discuss below, where the height of the peak is completely determined by the ``structure'' of the system while the peaks does not depend on the rate. For further details see App.~\ref{sec:two-continua}. In other words, the heights of the peaks can be tuned by the rates $\gamma_i$, and the revival times are given by $\tau_i$ \footnote{This discussion assumes that the revival peaks are completely suppressed when the next revival time sets in. In the opposite case on should expect further interference effects between contributions from different revival-time intervals.}

In Fig.~\ref{fig:two-bands} we display an example. The upper panel shows the exponential decay into two different  continua at short times. The lower panel shows the redistribution-dynamics of probability at long timescales, first from the ``blue'' continuum to the ``black,'' since the blue bans has the shorter revival time. The transfer of probability takes place via the initial state manifesting itself as a transient revival of that state accompanying each partial revival.  In the figure the first (blue) revival takes place around $t = 61$ au, the second (black) around $t = 102$ au. At $t=2 \times 61$ au we see a weak signature in the initial state of the second blue rival time and then around $t=161$ we observe a strong new revival at $t = \tau_1 + \tau2$, as expected. At increasing integer number of linear combinations of the two isolated revival times the two continua interchange probability each time, and each interchange imply, on this time-scale and coupling parameters, a spike-like re-population of the initial state.

Both methods can be directly expanded to describe more than two parallel bands, which in general will display
even more complex dynamics as long as the revival times of each band differ.

\section{Harmonic coupling}
\label{sec:harmonic}

Another class of atom-field couplings are coupling parameters which are periodic. The periodicity can originate from partial reflection of the radiation emitted from an excited atom, where the partial reflected part at some later time interfere with a later emitted radiation which is unreflected \cite{dorner2002laser}. Another example is a qubit connected to a substrate at two points separated a distance $2L$. The coupling from the two-level system to the surface plasmon states were recently analyzed in detail \cite{guo2017giant}. Writing $\beta_n = \beta \cos \big(n \Delta T/2\big)$, in the continuum limit one derived the elegant solution, which in our notations reads
\begin{equation}
\label{eq:harmonic-cont}
a(t) = \sum_{n=0}^\infty \frac{\left[-\frac{\gamma}4 (t-nT) \right]^n}{n!} \rme^{-(t-n T)\gamma/4} \Theta\big(t-nT\big) \,,
\end{equation}
where we again have chosen $a(0)=1$ and neglected the detuning energy.  This structure is considerable simpler than the one for a constant coupling, see Eqs.~\eqref{eq:const-sol-1} and \eqref{eq:ak-final}. In particular, note the absence of multiple peaks at later revival times $t > n T$ \footnote{This becomes clear from the simpler polynomial structure accompanying the exponentials in the sum.}. In order to address this ``simplicity'', we set out to analyze specific cases when the band density, concretely $2\pi/\Delta$, are multiples of the modulation time-scale $T$. 

We now address the dynamics using the discrete basis eigenvector approach. Consider Eq.~\eqref{eq:H} and replace the constant coupling parameter $\beta$ with an energy dependent coupling $\beta_n = \beta \cos (n \Delta T/ 2)$. In the analysis in \cite{guo2017giant}, the period $T$ is related to the the distance $L$ between two contact points between the qubit and the substrate,  $T= L/v_g$ where $v_g$ is the speed of sound in the acoustic medium. By defining a symmetric and antisymmetric discrete phonon states from the basis applied in \cite{guo2017giant}, only the symmetric ones couple to the qubit and we arrive at our Eq.~\eqref{eq:H}. The eigenstates are then given by the sum of Eq.~\eqref{eq:implicit}. Let us rewrite the sum in terms of 
\begin{equation}
{\cal S} = \sum_{n=-\infty}^\infty \frac{\left|\beta \cos \big(n \Delta T/2 \big) \right|^2}{E - \Delta n} \,.
\end{equation}
A general solution can be found using special functions. Here, we consider a few illustrative cases, using increasingly smaller energy separation between the discrete states:

\paragraph{$\Delta = 2\pi/T$:} This is equivalent to a constant coupling $\beta_n=\beta$ to all the energy levels. The answer is 
\begin{equation}
\label{eq:sum_harm1}
\frac{\Delta}{\beta^2}{\cal S} = \frac\pi{\tan \left[ \pi z\right]} \,,
\end{equation}
where $z = E/\Delta$. Further discussion can be found in Sec.~\ref{sec:single-continuum}.

\paragraph{$\Delta = \pi/T$:} The constant coupling $\beta$ is modulated by the following possible values $\cos(\omega_n T/2) = \{1,0,-1,\ldots \}$ for $n=\{0,1,2,\ldots\}$ et cetera. The sum therefore becomes
\begin{align}
\label{eq:sum_harm2}
\frac{\Delta}{\beta^2} {\cal S} = \sum_{n=-\infty}^\infty \frac{1}{z-2n} = \frac{\pi}{2 \, \tan \left[\frac{\pi}{2}z \right]}
\end{align}
This is equivalent to a constant coupling $\beta_n=\beta$ to all the energy levels.
This choice of $\Delta$ gives the correct exponential decay up to $t=T$ with a decay rate constant half the magnitude of the decay constant of Eq.~\eqref{eq:dyn_a1}, namely $\gamma' =  \frac{\pi\beta^2}{\Delta} = \gamma/2$. Further discussion can be found in App.~\ref{sec:app-harm2-coupling}.

At $t=T$ the back-coupling to the initial state departs from the correct continuum behavior. From Eq.~\eqref{eq:a1-constant}, we find a first peak at $t_{1} = T + 4/\gamma$ reaching $P(t_{1}) = 4/\rme^2$. This is a factor 4 too large compared to the continuum solution from Eq.~\eqref{eq:harmonic-cont}. Obviously, the multiple peaks associated with revivals at $t > 2 T$, discussed in detail in Sec.~\ref{sec:single-continuum}, are also not present in the continuum limit.

\paragraph{$\Delta = 2\pi/(3T)$:} The constant coupling $\beta$ is now modulated by the following possible values $\cos(\omega_n T/2) = \{1,\frac12,-\frac12,\ldots \}$ for $n=\{0,1,2,\ldots\}$. The sum therefore becomes
\begin{align}
\label{eq:sum_harm3}
\frac{\Delta}{\beta^2} {\cal S} &= \sum_{n=-\infty}^\infty \frac{1}{z-3n} \nonumber\\
& +\frac14 \sum_{n=-\infty}^\infty \left[ \frac{1}{z-(3n-1)} + \frac{1}{z-(3n+1)} \right]\,,\nonumber\\
&= \frac{\pi}{3 \, \tan\left[\frac\pi3 z \right]} - \frac\pi3 \frac{\sin\left[\frac{2 \pi}{3}z  \right]}{1+ 2 \cos \left[\frac{2\pi}{3}z \right]} \,.
\end{align}
This representation describes the dynamics up to the revival time $t=2T$. The couplings and energy levels of of the discrete states distribute themselves in what can be viewed as to parallel continua:  The first one has a the coupling strength and energy separation $|\beta,3\Delta|$, the second  $|\beta/2,\Delta|$, in addition making sure that the overlapping levels are not occupied twice. 

We have worked out this case in detail up to times $t \leq 3T$ in App.~\ref{sec:app-harm3-coupling}. For the time-interval $0 < t < 2T$, we find that 
\begin{equation}
a(t) = \rme^{- t\gamma/4} - \frac\gamma4 (t-T)\rme^{-(t-T)\gamma/4} \Theta(t-T) \,.
\end{equation}
At the first revival at $t=T$, the peak position remains at $T + 4/\gamma$, but now the height is reduced to $P= 1/\rme^2$, which is in agreement with the continuum result in Eq.~\eqref{eq:harmonic-cont}. Strikingly, as for the case of constant coupling to a single (quasi-) continuous band in Sec.~\ref{sec:single-continuum}, the height of the first peak is completely independent of the coupling or level density spacing. This also holds for the subsequent peaks. Only their positions are occurring at intervals $nT + c/\gamma$, where $c$ is number describing the positions of the multiple peaks. However, at $t\geq 3T$ this model starts exhibiting multiple peaks and, thus, does not agree anymore with the continuum solution \eqref{eq:harmonic-cont}, so we will not discuss it further here. 

\paragraph{$\Delta = \pi/(2T)$:} The constant coupling $\beta$ is now modulated by the following possible values u$\cos(\omega_n T/2) = \{1,\frac{1}{\sqrt{2}},0,\ldots \}$ for $n=\{0,1,2,\ldots\}$, et cetera. The sum therefore becomes
\begin{align}
\label{eq:sum_harm4}
\frac{\Delta}{\beta^2} {\cal S} &= \sum_{n=-\infty}^\infty \frac{1}{z-4n} \nonumber\\
&+\frac12 \sum_{n=-\infty}^\infty \left[ \frac{1}{z-(4n-1)} + \frac{1}{z-(4n+1)} \right]\,,\nonumber\\
&= \frac{\pi}{4 \tan \left[\frac\pi4 z \right]} - \frac\pi4 \tan\left[ \frac\pi2 z\right] \,.
\end{align}
 Again this is effectively two parallel continua, avoiding any overlaps, now with parameters $|\beta,4\Delta|$ and $|\beta/2,\Delta|$. We have worked this case out in detail in App.~\ref{sec:app-harm4-coupling}.
 
 This particular discretization gives identical population in the two continua up to $t=T$ and the redistribution  gives the correct dynamics of the initial state up to $t=3T$. Since up to $t=2T$ the result is exactly the same as in the previous cases, let us only write out the additional piece that appears at $t>2T$, namely
\begin{equation}
a^{(2)}(t) = \frac{\gamma^2}{32} (t-2T)^2 \rme^{-(t-2T)\gamma/4} \Theta(t-2T) \,.
\end{equation}
 Now, there is only one peak in the interval $2T < t < 3T$, located at $t_2= 2 T+\frac8\gamma$, where the probability peaks at $P(t_2) = 4/\rme^4\approx 0.073$. The agreement with the continuum result now extends to $t < 3T$. Again, the height of the second peak does not depend on the parameters of the model.

\begin{figure}[t] 
\begin{center}
\includegraphics[width=\columnwidth]{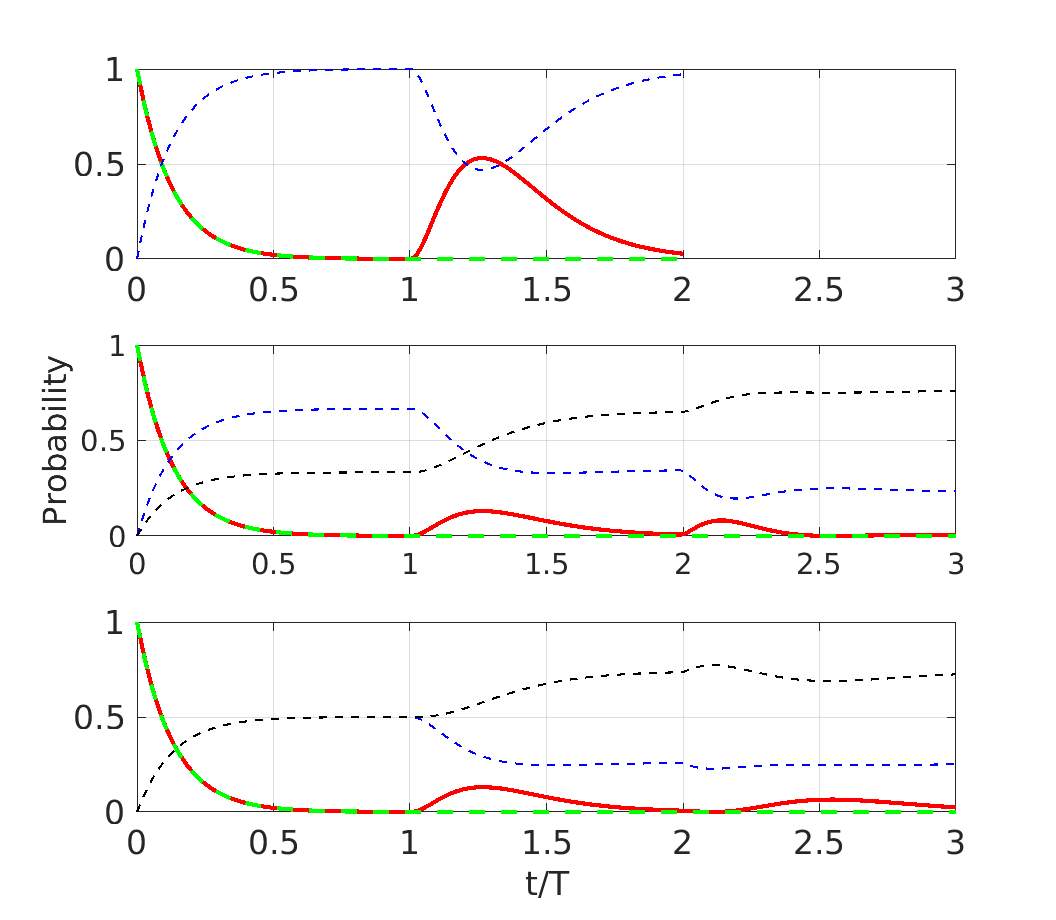}
\caption{Dynamics of the initial qubit state (red) and parallel "phonon continua" (dotted blue,black curves)
for varying energy separation, $\Delta=2\pi/(M\cdot T)$. Upper panel: M=2, middle panel: M=3 and lower panel: M=4. The thick broken curved line is the analytical exponential decay. The red curve in the lowest panel is identical with the exact result, cf. \cite{guo2017giant}.}
\label{fig:harmonic-coupling}
\end{center}
\end{figure}

In summary, the dynamics of the qubit with harmonic coupling to a substrate can alternative be viewed as the dynamics of a series of parallel continua with different constant coupling and density of states. At least one parallel continuum has a revival at integer number of $T$ which causes back-coupling to the initial state followed by a redistribution of the population probability of each quasi-continuum. The dynamics described above is illustrated in Fig.~\ref{fig:harmonic-coupling}.

\begin{figure}[t] 
\begin{center}
\includegraphics[width=\columnwidth]{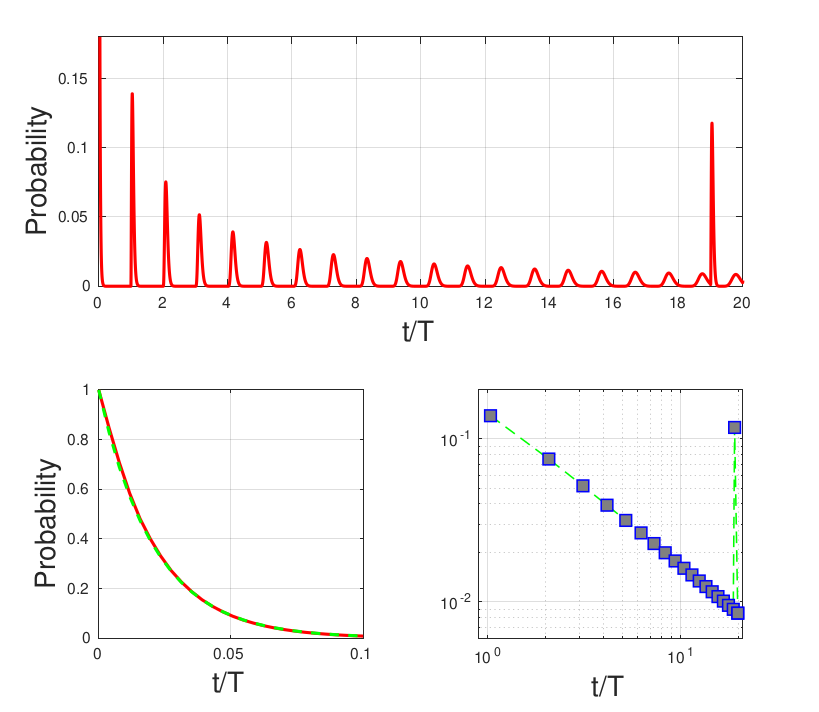}
\caption{The upper panel shows the probability of the initial state in a simulation with 7000 states in total. Parameters: $T=1, \ \Delta= 2\pi/(M*T), \ M=20$ and $\beta = 2.1859$. The lower left panel shows the simulated and analytical initial time developments of the initial state probability and the lower right panel displays the peak values of the initial state revivals.}
\label{fig:harmonic-late}
\end{center}
\end{figure}

By taking a sufficiently small $\Delta$ we can effectively simulate the dynamics for arbitrarily large times. In Fig.~\ref{fig:harmonic-late} this is done for a case which has $\tau\approx  19 T$. A large number of initial state revivals is seen up to the point $t=\tau$ where \emph{all} parallel continua has a back-coupling to the initial state. As a consequence this peak is much larger than the partial revivals at integer values of $t/T$. In the lower left panels the initial decay at very short time again is observed, and in full agreement with the analytical decay rate constant. Finally, as also shown in the case of constant coupling, the revival peaks display a polynomial decay up to $t=\tau$.

\begin{figure}[ht] 
\begin{center}
\includegraphics[width=\columnwidth]{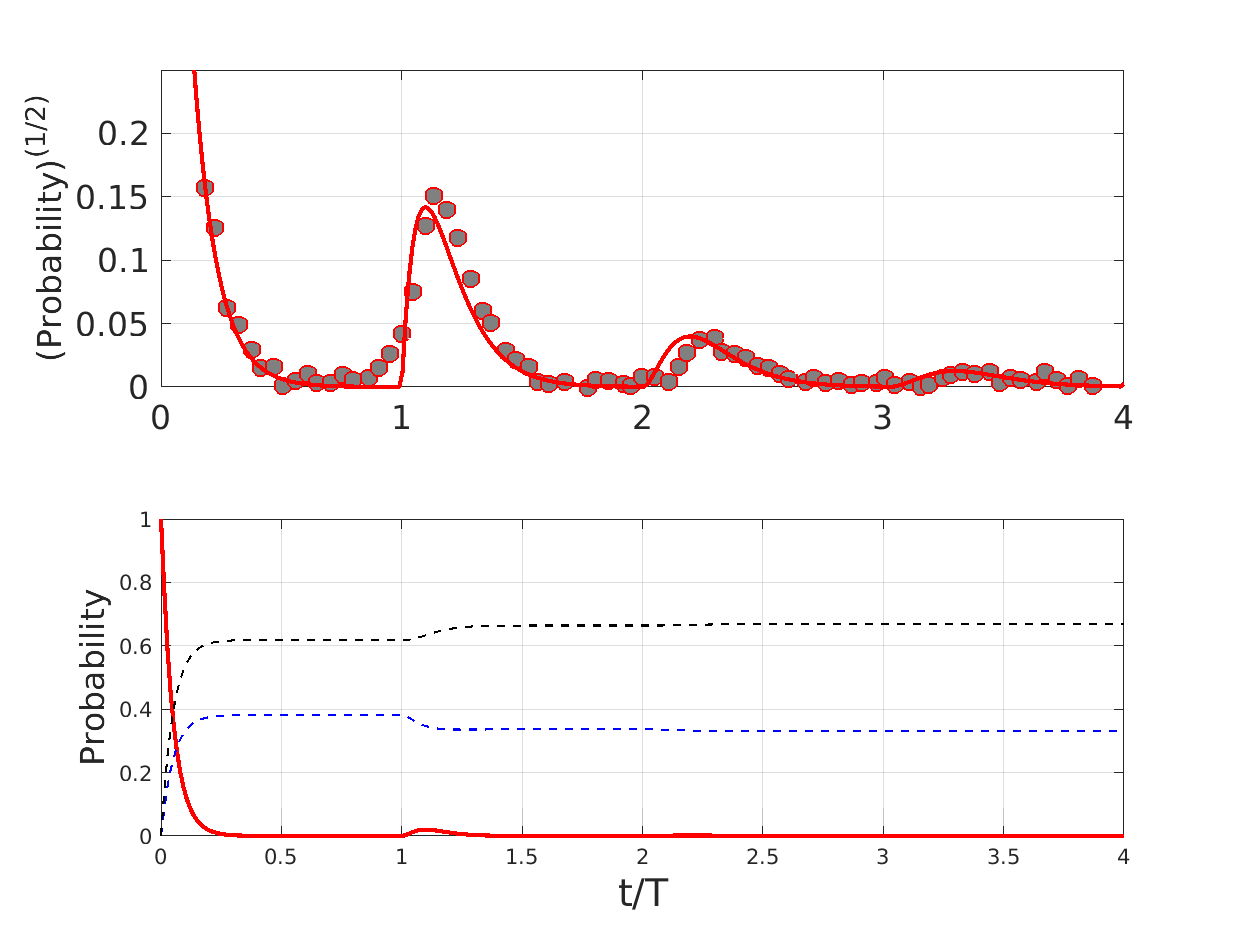}
\caption{
The upper panel shows the absolute amplitude of the initial state in a simulation with 1000 states compared with the experimental results (circles) in \cite{andersson2019non}. Parameters: $T=4, \ \Delta= 2\pi/(M*T), \ M=20$ and $\beta = 0.4372$. In addition a ``heat reservoir'' described by 5000 states and a factor five denser energy separation, $\Delta_{heat}=\Delta/5$, has been added. The coupling from the excited qubit state to the heat reservoir is $\beta_{heat}=0.176$. The lower panel shows the time dependent probability distributions of the excited qubit state (red thick line), summed phonon (blue broken line) and reservoir state (black broken line).}
\label{fig:experimental}
\end{center}
\end{figure}

The use of discrete parallel continua makes it simple to model experiments as well, for example in the case of 
Ref.~\cite{andersson2019non} where a number of other couplings appear due to temperature fluctuations or other conditions related to the experiment. By simply augmenting the discrete model ($M=4$) with a dense reservoir which on the timescale of the experiment does not undergo any revivals we can simulate a leak of probability due to heat.  The result, shown in Fig.~\ref{fig:experimental} is in excellent agreement with experimental data from \cite{andersson2019non}. 

\section{Concluding remarks}
\label{sec:conclusions}

I this work we have considered the dynamics of a quantum model which goes back almost to the invention of quantum mechanics. Inspired by the recent progress and application of the model relevant for qubit dynamics of modern quantum computers, we analyzed the general problem based on a single state interacting with one or several discrete band(s). The analysis were performed in an eigenvector basis and with a complementary analytical method applicable to real as well as discrete ``continua''. 
A number of properties of known and new phenomena were identified, in particular:
\begin{enumerate}
\item[(i)] Discrete band of states imply well defined revival peaks. For a single band the heights of the 
peaks are independent of coupling strengths and the peak strengths shows near polynomial decays on a time interval covering several tens of revival times.
\item[(ii)] The qubit dynamics with harmonic couplings between the initial state and a true (phonon) continuum is identical to a qubit coupled by constant  factors to a number of parallel bands. Revivals in the latter model are understood as a transient population flow through the initial state and between different bands.
\item[(iii)] Comparison with experiments based on the numerical approach was straightforward and in excellent  overall agreement.
\end{enumerate}
The present approach opens for direct modeling of time dependent scattering phenomena due to point contacts, of time induced multi-phonon (photon) processes and a number of other phenomena relevant for the long research path towards the realization of stable and effective working quantum computers.

\begin{acknowledgments}
One of us (JPH) would like to thank Ladislav Kocbach and Lars Bojer Madsen for initial inspiring discussions 
on the topic of this work.
\end{acknowledgments}

\appendix
\section{Laplace transform method}
\label{sec:laplace}

The quantum evolution described above can also be cast as a system of evolution equations for the state $a(t)$ coupled to a quasi-continuum $b_n(t)$,
\begin{align}
\dot a(t) & = - i \sum_{n=-\infty}^\infty \beta_n b_n(t) \,,\\
\dot b_n(t) &= -i \omega_n b_n(t) - i \beta^\ast_n a(t) \,,
\end{align}
where $\omega_n = n \Delta$, and $\Delta$ is the level spacing of the quasi-continuum. Integrating out the second equation, yielding
\begin{equation}
\label{eq:bn-solution}
b_n(t) = -i \beta_n \int_0^t \rmd t' \, \rme^{-i \omega_n(t-t')}a(t') \,,
\end{equation}
and inserting the solution into the first, we obtain the standard solution
\begin{equation}
\label{eq:an-solution}
\dot a(t) = - \int_0^t {\rm d}t' \, a(t') \sum_{n= -\infty}^\infty |\beta_n|^2 {\rm e}^{- i \omega_n(t-t')} \,.
\end{equation}
This has the form of a delay differential equation \cite{milonni1983exponential}.

This can be solved in Laplace space, where the Laplace transform of $a(t)$, defined as
\begin{equation}
\tilde a(s) = \int_0^\infty {\rm d} t \, {\rm e}^{-s t} a(t) \,,
\end{equation}
is given by
\begin{equation}
\tilde a(s) = \frac{a(0)}{s + \Pi(s) } \,,
\end{equation}
where $\Pi(s) = i\sum_{n=-\infty}^\infty \frac{|\beta_n|^2}{is - \omega_n} = i {\cal S}(is)$. Note that $\Pi(-s) = -\Pi(s)$.
The inverse transform, defined as
\begin{equation}
a(t) = \int_C \frac{{\rm d} s}{2\pi i} \,{\rm e}^{s t}\, \tilde a(s) \,,
\end{equation}
where the contour $C$ runs along the imaginary axis to the right of any poles of $\tilde a(s)$, then becomes sensitive to the poles of $\tilde a(s)$. These are found by solving
\begin{equation}
s + \Pi(s) =0 \,.
\end{equation}
By identifying $i s \leftrightarrow E$, the pole structure is equivalent to the eigenvalue equations.

\subsection{Constant coupling}
\label{sec:app-const-coupling}
As mentioned above, a constant coupling is equivalent to setting the level spacing $\Delta = 2\pi/T$ in the harmonic coupling. Using the sum derived in \eqref{eq:sum_harm1}, we find
\begin{equation}
\Pi(s) = i \frac\gamma2 \cot (i \ell s ) = \frac\gamma2 \coth(\ell s) \,,
\end{equation}
where $\gamma = 2\pi \beta^2/\Delta$ and $\ell = \pi/\Delta$. In order to introduce the techniques to inverse the Laplace transform \cite{stey1972decay}, we will review this case in  detail.

It is possible to rewrite,
\begin{equation}
\tilde a(s ) = \frac{1- \rme^{-2 \ell s}}{s + \gamma/2} \left[1 - \frac{s- \gamma/2}{s+\gamma/2}\rme^{-2 \ell s} \right]^{-1} \,.
\end{equation}
Since $|(s- \gamma/2)/(s+\gamma/2) \rme^{-2 \ell s}|< 1$ for sufficiently large $\ell$, or small $\Delta$, 
we can expand the last term as an infinite series $(1-x)^{-1} = \sum_{n=0}^\infty x^n$. Then, after further manipulations, we write the whole expressions as
\begin{equation}
\tilde a(s) = \frac{1}{s+\gamma/2} -\gamma \sum_{k=1}^\infty \frac{(s-\gamma/2)^{k-1}}{(s+\gamma/2)^{k+1}} \rme^{-2 k \ell s} \,.
\end{equation}
This demonstrates that there is only one location of the poles, i.e. at $s_\ast = - \gamma/2$. We can therefore safely let the contour $C$ run along the origin of the complex $s$ plane, i.e. $s = i \xi$ with $\xi \in [-\infty, \infty]$ with a counterclockwise circle at infinity. Defining 
\begin{equation}
a(t) = a_0(t) + \sum_{k=1}^\infty a_k(t) \,,
\end{equation}
where
\begin{equation}
a_0(t) = \int_C \frac{\rmd s}{2\pi i} \, \frac{\rme^{st} }{s+\gamma/2} = \rme^{- t \gamma/2 } \,,
\end{equation}
and 
\begin{equation}
a_k(t) = \int_C \frac{\rmd s}{2\pi i} \, \tilde a_k(s) \rme^{st} \,,
\end{equation}
with $\tilde a_k(s) = -\gamma \frac{(s-\gamma/2)^{k-1}}{(s+\gamma/2)^{k+1}}\rme^{-2 k\ell s}$.
The coefficients $a_k(t)$ can easily be found using the residue theorem,
\begin{equation}
a_k(t) =  \frac{1}{k!} \lim_{s\to s_\ast} \frac{\rmd^k}{{\rmd s}^k} \left[(s+\gamma/2)^{k+1}\, \tilde a_k(s) \rme^{st} \right] \,,
\end{equation}
as long as the function decays fast enough on the circle at infinity, in this case as long as $t > 2 k \ell s$. We use the binomial formula to write
\begin{multline}
\frac{\rmd^k}{{\rmd s}^k} \left[ (s-\gamma/2)^{k-1} \rme^{(t- 2 k\ell)s} \right] 
\\ = \sum_{n=0}^{k-1} \frac{(k-1)!}{(k-1-n)!}(s-\gamma/2)^{k-1-n} (t-2k \ell)^{k-n} \rme^{(t-2k\ell)s} \,.
\end{multline}
Finally, after further manipulations, we can recast the answer as
\begin{equation}
\label{eq:ak-final-2}
a_k(t) = - \frac{\gamma(t-k \tau)}{k}\rme^{-(t-k\tau)\gamma/2} L_{k-1}^{\{1\}}\big(\gamma(t-k \tau)\big) \Theta\big(t- k\tau \big) \,,
\end{equation}
where we defined $\tau = 2\pi/\Delta$ and $L_{n}^{\{\alpha\}}(x)$ is the generalized Laguerre function. Note, that in this case we have exactly that $\tau = T$. 

The occupation of the continuum states is found from Eq.~\eqref{eq:bn-solution}. For instance, up to the first revival time, i.e. $0 < t < \tau$, where $P_a(t) = \rme^{- t \gamma}$, the solution is
\begin{equation}
b_n(t) = \frac{i \beta}{\frac{\gamma}2 - i n \Delta} \left( \rme^{-t \gamma/2} - \rme^{-i n \Delta t} \right) \,.
\end{equation}
The probability for the whole continuum then becomes
\begin{equation}
P_b(t) = \sum_{n=-\infty}^\infty \left| b_n(t) \right|^2 \,.
\end{equation}
To prove that the total probability is conserved, we use the following identities
\begin{align}
\label{eq:sum-1}
\sum_{n=-\infty}^\infty\frac{1}{z^2 + n^2} &= \frac{\pi}{z}\coth\left( z \pi \right) \,,\\
\label{eq:sum-2}
\sum_{n=-\infty}^\infty\frac{\cos\left(x n \right)}{z^2 + n^2} &= \frac{\pi}{z}\frac{\cosh\left( z (\pi - x) \right)}{\sinh\left( z\pi \right)} \,,
\end{align}
for $0 < x < 2\pi$ \cite{Gradshteyn:1943cpj}, to derive that
\begin{equation}
P_b(t) = 1- \rme^{-t \gamma } \,.
\end{equation}
This demonstrates that $P_a(t) + P_b(t) = 1$, as expected.

\subsection{Harmonic coupling with $\Delta = \pi/T$}
\label{sec:app-harm2-coupling}

Using \eqref{eq:sum_harm2} to obtain, that
\begin{equation}
\Pi(s) = \frac\gamma4 \coth\left(\frac{\ell s}{2} \right) \,,
\end{equation}
we realize that the pole structure and therefore also the inverse Laplace transform is completely equivalent to the previous case with the replacement $\gamma \to \gamma/2$ and the explicit mapping of $\tau \to T$. The first correction therefore reads,
\begin{equation}
a_1(t) = -\frac\gamma2 \left(t-T\right)\rme^{-(t-T)\gamma/4} \Theta\big(t-T \big) \,,
\end{equation}
in this case. Further terms, appearing at $t>2T$, can be found directly from previous formulas.

\subsection{Harmonic coupling with $\Delta = 2\pi/(3T)$}
\label{sec:app-harm3-coupling}

In this case, with the help of \eqref{eq:sum_harm3}, we found
\begin{equation}
\Pi(s) = \frac\gamma6 \coth \left( \frac{\ell s}{3}\right) +\frac\gamma6 \frac{\sinh \left(\frac23 \ell s \right)}{1+2 \cosh \left(\frac23 \ell s \right)} \,.
\end{equation}
Following the same line of arguments as before, and introducing the short-hand $y \equiv \rme^{-\frac23 \ell s}$, we find that
\begin{equation}
\tilde a(s) = \frac{1-y^3}{s+ \gamma/4} \left[1 - \frac{(s-\gamma/4)y^3 - (y+y^2)\gamma/4}{s+\gamma/4} \right]^{-1} \,.
\end{equation}
After further manipulations, we arrive at
\begin{multline}
\label{eq:tildea-cos-twolevel}
\tilde a(s) = \frac{1}{s+\gamma/4} - \frac\gamma2 \sum_{k=1}^\infty \frac{ \left[(s-\gamma/4)y^3 - (y+y^2)\gamma/4 \right]^{k-1}}{(s+\gamma/4)^{k+1}} \\ \times \left(y^3 + y^2/2 + y/2\right) \,.
\end{multline}
We identify the emerging multiple poles at $s=s_\ast = - \gamma/4$, however the structure of the numerator is much more complicated. It is possible proceed further by expanding the numerator by using the binomial formula, but we will not proceed further.

Instead, let us first have a look at the first term in the expansion in \eqref{eq:tildea-cos-twolevel}, which reads
\begin{align}
\tilde a_1(s) 
&= -\frac\gamma2 \frac{\rme^{-3 T s} + \rme^{-2Ts}/2 + \rme^{-Ts}/2}{(s+\gamma/4)^2} \,.
\end{align}
The three terms in the denominator will contribute to the evolution only at $t>3T$, $t>2T$ and $t>T$, respectively. Similarly, $\tilde a_2(s)$ will contain terms that contribute at subsequent time intervals, starting at $t>2T$ and up to $t>6T$. We will currently only be interested in the terms that describe the evolution in then interval $0 < t< 3T$. The contribution that starts to contribute at times $n T < t$, we will denote $a^{(n)}(t)$. Naturally, $a^{(0)}(t) = \rme^{- t\gamma/4}$ from the single pole contribution.

Focussing now on the next interval, $T< t<2T$, we note that only $\tilde a_1(s)$ contributes. We find
\begin{align}
a^{(1)}(t) &= -\frac\gamma4 \int_C \frac{\rmd s}{2\pi i} \, \frac{\rme^{(t-T)s}}{(s+\gamma/4)^2} \nonumber\\
&=-\frac\gamma4 (t-T)\rme^{-(t-T)\gamma/4} \Theta(t-T) \,.
\end{align}
This piece reaches a maximal value at $t = T+2/\gamma$, as before, where the contribution to the probability reaches a maximal value of $\rme^{-2}\approx 0.14$. This is a factor 4 smaller than in the previous cases!

The additional contribution in the next piece, i.e. $t > 2T$, receives contributions from both $\tilde a_1(s)$ and $\tilde a_2(s)$. We find
\begin{align}
a^{(2)}(t) &= -\frac\gamma4 \int_C\frac{\rmd s}{2\pi i} \,\frac{\rme^{(t-2T)s}}{(s+ \gamma/4)^2} + \frac{\gamma^2}{16} \int_C \frac{\rmd s}{2 \pi i} \,\frac{\rme^{(t-2T)s}}{(s+ \gamma/4)^3} \,,\nn
&= \left[\left(-\frac\gamma4 \right)(t-2T) + \frac{\gamma^2}{32}(t-2T)^2\right] \nonumber\\
&\times \rme^{-(t-2T)\gamma/4} \Theta(t-2T) \,.
\end{align}
There are two maxima at $t = 2T +\frac4\gamma(2\mp \sqrt2)$,
corresponding to peak values on the level of the probability that are $\{(3-2\sqrt2)\rme^{-4+2\sqrt2}, \, (3+2\sqrt2)\rme^{-4-2\sqrt2} \} \approx \{0.053 , 0.006\}$.

Compared to the continuum limit, $\Delta \to 0$, we find agreement of the height of the peak in the interval $T < t < 2T$. The multiple peaks occurring at later times are also not present in the $\Delta \to 0$ solution.

\subsection{Harmonic coupling with $\Delta = \pi/(2T)$}
\label{sec:app-harm4-coupling}

In this case, from Eq.~\eqref{eq:sum_harm4}, we have
\begin{equation}
\Pi(s) = \frac\gamma8 \coth \left(\frac{\ell s}{4} \right) + \frac\gamma8 \tanh\left(\frac{\ell s}{2}\right) \,.
\end{equation}
After standard manipulations, we arrive at
\begin{multline}
\tilde a(s) = \frac{1}{s+\gamma/4} - \frac\gamma2 \sum_{k=1}^\infty \frac{ \left[(s-\gamma/4)y^3 + (y-y^2) s \right]^{k-1}}{(s+\gamma/4)^{k+1}} \\ \times \left(y^3 - y^2/2 + y/2\right)\,.
\end{multline}
Proceeding as before, we find the following behavior for the early time evolution,
\begin{align}
a^{(1)}(t) &= -\frac\gamma4 (t-T) \rme^{-(t-T)\gamma/4} \Theta(t-T) \,,\\
a^{(2)}(t) &= \frac{\gamma^2}{32} (t-2T)^2 \rme^{-(t-2T)\gamma/4} \Theta(t-2T) \,.
\end{align}
Now, there is only one peak in the interval $2T < t < 3T$, located at $t= 2 T+\frac8\gamma$, where the probability peaks at $4/\rme^4\approx 0.073$. The agreement with the continuum result now extends to $t < 3T$. 

We end with a few comments. It becomes evident that the late time behavior of the system is highly sensitive to the spacing of the energy levels. The peaks occurring at later time intervals are also increasingly shifted upward. For $\Delta > 0$, at some late time the subtle cancellations do not hold any longer and violent, chaotic behavior can occur due to the overlap of several contributions. 

\section{Coupling to two quasi-continua}
\label{sec:two-continua}

The evolution equations for a coupling to two quasi-continua $b_n(t)$ and $c_m(t)$ read,
\begin{align}
\dot a(t) & = - i \sum_{n=-\infty}^\infty \beta^{(1)}_n b_n(t) - i \sum_{m=-\infty}^\infty \beta^{(2)}_m c_m(t) \,,\\
\dot b_n(t) &= -i \omega^{(1)}_n b_n(t) - i \beta^{(1),\ast}_n a(t) \,,\\
\dot c_m(t) &= -i \omega^{(2)}_m c_m(t) - i \beta^{(2),\ast}_m a(t) \,,
\end{align}
where the two energy levels are give by $\omega^{(i)}_n = n \Delta_i$ ($i=1,2$). The solution in Laplace space is given by
\begin{equation}
\tilde a(s) = \frac{a(0)}{s + \Pi_1(s) + \Pi_2(s) } \,,
\end{equation}
where $\Pi_j(s) = i \sum_{n=-\infty}^\infty |\beta^{(j)}_n|^2/ (is - \omega^{(j)}_n)$.
We focus on the case of constant couplings, i.e. $\beta^{(i)}_n = \beta_i$. After, by now, standard manipulations we finally arrive at
\begin{align}
&\tilde a(s) = \frac1{s+\frac{\gonetwo}2} \nonumber\\
&- \sum_{k=1}^\infty \frac{\left[\left(s - \frac{\gamma_1}2 \right)\rme^{-\tau_1 s} + \left(s - \frac{\gamma_2}2 \right)\rme^{-\tau_2 s} - s \rme^{-\tau_{12}s} \right]^{k-1}}{\left( s+\frac{\gonetwo}2\right)^{k+1}} \nonumber\\
&\times \left[ \left(\gamma_1 +\frac{\gamma_2}2 \right) \rme^{-\tau_1 s} + \left(\gamma_2 +\frac{\gamma_1}2 \right) \rme^{-\tau_2 s} - \frac{\gonetwo}2\rme^{-\tau_{12}s} \right] \,,
\end{align}
where $\gamma_i = 2\pi |\beta_i|^2/\Delta_i$ and $\tau_i = 2\pi/\Delta_i$ and, finally, $\gonetwo = \gamma_1 + \gamma_2$ and $\tau_{12} = \tau_1+ \tau_2$. 
The amplitude picks again up new contributions at multiples of the revival times, and also allows arbitrary mixing between multiples of  $\tau_1$ and $\tau_2$, e.g. $n\tau_1 \times m \tau_2$, and so forth. The first few terms in the expansion are
\begin{align}
a_0(t) &= \rme^{- t \gamma_{12}/2} \,,\\
a_1(t) &= -\left(\gamma_1+\frac{\gamma_2}2 \right) (t-\tau_1) \rme^{-(t-\tau_1)\gonetwo/2} \Theta(t-\tau_1)\nonumber\\
&-\left(\gamma_2+\frac{\gamma_1}2 \right) (t-\tau_2) \rme^{-(t-\tau_2)\gonetwo/2} \Theta(t-\tau_2) \nonumber \\
&+\frac{\gonetwo}2 (t- \tau_{12}) \rme^{-(t-\tau_{12}) \gonetwo/2} \Theta(t-\tau_{12}) \,,
\end{align}
At second order, we get contributions at revival times $t > 2 \tau_1,\, 2 \tau_2,\, \tau_{12}$ and also $t > 2 \tau_1+\tau_2,\, \tau_1 + 2\tau_2, \, 2 \tau_{12}$. The former contributions are fully accounted by the two first terms, while the latter will receive further contributions from the third term of the expansion. Let us therefore focus on the former for now. The contributions read
\begin{align}
a_3(t) &= -\left(\gamma_1+\frac{\gamma_2}2\right)\left[(t-2\tau_1) -\frac12\left(\gamma_1+\frac{\gamma_2}2\right)(t-2\tau_1)^2 \right] \nonumber\\
&\qquad \times \rme^{-(t-2\tau_1)\gonetwo/2} \Theta(t-2\tau_1) \nonumber\\
& -\left(\gamma_2+\frac{\gamma_1}2\right)\left[(t-2\tau_2) -\frac12\left(\gamma_2+\frac{\gamma_1}2\right)(t-2\tau_2)^2 \right] \nonumber\\
&\qquad \times \rme^{-(t-2\tau_2)\gonetwo/2} \Theta(t-2\tau_2) \nonumber\\
&- \left[\frac{3 \gonetwo}{2}(t-\tau_{12}) -\left(\gamma_1+\frac{\gamma_2}2\right)\left(\gamma_2+\frac{\gamma_1}2\right) (t-\tau_{12})^2 \right]\nonumber\\
&\qquad \times \rme^{-(t-\tau_{12}) \gonetwo/2} \Theta(t-\tau_{12}) \,.
\end{align}

There are several distinct features that are worth pointing out. We already

The first peaks related to the revival times $\tau_i$ appear at times $t_i = \tau_i+2/\gonetwo$ where the probability peaks at $\left|\frac{2 \gamma_i + \gamma_j}{\rme\gonetwo} \right|^2$ (where $\gamma_j$ refers to the other coupling, i.e. $i\neq j$), respectively. Strikingly, the height of these peaks depends on the couplings. 

The revival time at $\tau_1+\tau_2$ is novel, and involves two peaks. The position and height of the peak can be found in a straightforward manner, but the expressions are complicated and not illuminating. Instead, focussing on the case when $\gamma_1=\gamma_2=\gamma$, we find that the revival times are located at $t = \tau_1+\tau_2 + (13\pm \sqrt{97})/(9 \gamma)$ where the probability peaks at $\left|\frac{9 \pm \sqrt{97}}{2} \rme^{-(13 \pm \sqrt{97})/9} \right|^2$. Strikingly, this peak is again \emph{not dependent} on the coupling. However, this statement does not hold for the general case when $\gamma_1 \neq \gamma_2$. We also repeat that these estimates only hold if the couplings are sufficiently large so that the previous revival peaks are completely washed out before the subsequent revival time. 

As usual for constant coupling, double peaks also appear at multiples of the revival times $\tau_1$ and $\tau_2$, and also at their mixed multiples.

Finally, we can also find the occupation of the two bands, using Eq.~\eqref{eq:bn-solution}. In our case,
\begin{equation}
b_n(t) = \frac{i \beta_1}{\frac{\gonetwo}2 - i n \Delta_1} \left(\rme^{- t\gonetwo/2} - \rme^{-i n \Delta_1 t} \right)\,,
\end{equation}
and similarly for $c_n(t)$ by substituting $\beta_1 \to \beta_2$ and $\Delta_1 \to \Delta_2$. We define the probabilities,
\begin{align}
P_b(t) = \sum_{n=-\infty}^\infty \left| b_n(t)\right|^2 \,,\\
P_c(t) = \sum_{n=-\infty}^\infty \left| c_n(t)\right|^2 \,,
\end{align}
and after some algebra, using Eqs.~\eqref{eq:sum-1} and \eqref{eq:sum-2}, we obtain Eq.~\eqref{eq:band-population}, which proves that $P_a(t) + P_b(t) + P_c(t) = 1$.

\bibliographystyle{apsrev4-2}
\bibliography{exp_decay.bib}

\begin{thebibliography}{24}%
\makeatletter
\providecommand \@ifxundefined [1]{%
 \@ifx{#1\undefined}
}%
\providecommand \@ifnum [1]{%
 \ifnum #1\expandafter \@firstoftwo
 \else \expandafter \@secondoftwo
 \fi
}%
\providecommand \@ifx [1]{%
 \ifx #1\expandafter \@firstoftwo
 \else \expandafter \@secondoftwo
 \fi
}%
\providecommand \natexlab [1]{#1}%
\providecommand \enquote  [1]{``#1''}%
\providecommand \bibnamefont  [1]{#1}%
\providecommand \bibfnamefont [1]{#1}%
\providecommand \citenamefont [1]{#1}%
\providecommand \href@noop [0]{\@secondoftwo}%
\providecommand \href [0]{\begingroup \@sanitize@url \@href}%
\providecommand \@href[1]{\@@startlink{#1}\@@href}%
\providecommand \@@href[1]{\endgroup#1\@@endlink}%
\providecommand \@sanitize@url [0]{\catcode `\\12\catcode `\$12\catcode
  `\&12\catcode `\#12\catcode `\^12\catcode `\_12\catcode `\%12\relax}%
\providecommand \@@startlink[1]{}%
\providecommand \@@endlink[0]{}%
\providecommand \url  [0]{\begingroup\@sanitize@url \@url }%
\providecommand \@url [1]{\endgroup\@href {#1}{\urlprefix }}%
\providecommand \urlprefix  [0]{URL }%
\providecommand \Eprint [0]{\href }%
\providecommand \doibase [0]{https://doi.org/}%
\providecommand \selectlanguage [0]{\@gobble}%
\providecommand \bibinfo  [0]{\@secondoftwo}%
\providecommand \bibfield  [0]{\@secondoftwo}%
\providecommand \translation [1]{[#1]}%
\providecommand \BibitemOpen [0]{}%
\providecommand \bibitemStop [0]{}%
\providecommand \bibitemNoStop [0]{.\EOS\space}%
\providecommand \EOS [0]{\spacefactor3000\relax}%
\providecommand \BibitemShut  [1]{\csname bibitem#1\endcsname}%
\let\auto@bib@innerbib\@empty
\bibitem [{\citenamefont {Dirac}(1927)}]{dirac1927quantum}%
  \BibitemOpen
  \bibfield  {author} {\bibinfo {author} {\bibfnamefont {P.~A.~M.}\
  \bibnamefont {Dirac}},\ }\href@noop {} {\bibfield  {journal} {\bibinfo
  {journal} {Proceedings of the Royal Society of London. Series A, Containing
  Papers of a Mathematical and Physical Character}\ }\textbf {\bibinfo {volume}
  {114}},\ \bibinfo {pages} {243} (\bibinfo {year} {1927})}\BibitemShut
  {NoStop}%
\bibitem [{\citenamefont {Landau}(1927)}]{landau1927dampfungsproblem}%
  \BibitemOpen
  \bibfield  {author} {\bibinfo {author} {\bibfnamefont {L.}~\bibnamefont
  {Landau}},\ }\href@noop {} {\bibfield  {journal} {\bibinfo  {journal}
  {Zeitschrift f{\"u}r Physik}\ }\textbf {\bibinfo {volume} {45}},\ \bibinfo
  {pages} {430} (\bibinfo {year} {1927})}\BibitemShut {NoStop}%
\bibitem [{\citenamefont {Weisskopf}\ and\ \citenamefont
  {Wigner}(1930)}]{weisskopf1930calculation}%
  \BibitemOpen
  \bibfield  {author} {\bibinfo {author} {\bibfnamefont {V.~F.}\ \bibnamefont
  {Weisskopf}}\ and\ \bibinfo {author} {\bibfnamefont {E.~P.}\ \bibnamefont
  {Wigner}},\ }\href@noop {} {\bibfield  {journal} {\bibinfo  {journal} {Z.
  Phys.}\ }\textbf {\bibinfo {volume} {63}},\ \bibinfo {pages} {54} (\bibinfo
  {year} {1930})}\BibitemShut {NoStop}%
\bibitem [{\citenamefont {Wilson}(1968)}]{wilson1968fermi}%
  \BibitemOpen
  \bibfield  {author} {\bibinfo {author} {\bibfnamefont {F.~L.}\ \bibnamefont
  {Wilson}},\ }\href@noop {} {\bibfield  {journal} {\bibinfo  {journal}
  {American Journal of Physics}\ }\textbf {\bibinfo {volume} {36}},\ \bibinfo
  {pages} {1150} (\bibinfo {year} {1968})}\BibitemShut {NoStop}%
\bibitem [{\citenamefont {Merzbacher}(1998)}]{merzbacher1998quantum}%
  \BibitemOpen
  \bibfield  {author} {\bibinfo {author} {\bibfnamefont {E.}~\bibnamefont
  {Merzbacher}},\ }\href@noop {} {\emph {\bibinfo {title} {Quantum
  mechanics}}}\ (\bibinfo  {publisher} {John Wiley \& Sons},\ \bibinfo {year}
  {1998})\BibitemShut {NoStop}%
\bibitem [{\citenamefont {Rothe}\ \emph {et~al.}(2006)\citenamefont {Rothe},
  \citenamefont {Hintschich},\ and\ \citenamefont
  {Monkman}}]{rothe2006violation}%
  \BibitemOpen
  \bibfield  {author} {\bibinfo {author} {\bibfnamefont {C.}~\bibnamefont
  {Rothe}}, \bibinfo {author} {\bibfnamefont {S.}~\bibnamefont {Hintschich}},\
  and\ \bibinfo {author} {\bibfnamefont {A.}~\bibnamefont {Monkman}},\
  }\href@noop {} {\bibfield  {journal} {\bibinfo  {journal} {Physical review
  letters}\ }\textbf {\bibinfo {volume} {96}},\ \bibinfo {pages} {163601}
  (\bibinfo {year} {2006})}\BibitemShut {NoStop}%
\bibitem [{\citenamefont {Peshkin}\ \emph {et~al.}(2014)\citenamefont
  {Peshkin}, \citenamefont {Volya},\ and\ \citenamefont
  {Zelevinsky}}]{peshkin2014non}%
  \BibitemOpen
  \bibfield  {author} {\bibinfo {author} {\bibfnamefont {M.}~\bibnamefont
  {Peshkin}}, \bibinfo {author} {\bibfnamefont {A.}~\bibnamefont {Volya}},\
  and\ \bibinfo {author} {\bibfnamefont {V.}~\bibnamefont {Zelevinsky}},\
  }\href@noop {} {\bibfield  {journal} {\bibinfo  {journal} {Europhysics
  Letters}\ }\textbf {\bibinfo {volume} {107}},\ \bibinfo {pages} {40001}
  (\bibinfo {year} {2014})}\BibitemShut {NoStop}%
\bibitem [{\citenamefont {Kelkar}\ \emph {et~al.}(2004)\citenamefont {Kelkar},
  \citenamefont {Nowakowski},\ and\ \citenamefont
  {Khemchandani}}]{kelkar2004hidden}%
  \BibitemOpen
  \bibfield  {author} {\bibinfo {author} {\bibfnamefont {N.}~\bibnamefont
  {Kelkar}}, \bibinfo {author} {\bibfnamefont {M.}~\bibnamefont {Nowakowski}},\
  and\ \bibinfo {author} {\bibfnamefont {K.}~\bibnamefont {Khemchandani}},\
  }\href@noop {} {\bibfield  {journal} {\bibinfo  {journal} {Physical Review
  C}\ }\textbf {\bibinfo {volume} {70}},\ \bibinfo {pages} {024601} (\bibinfo
  {year} {2004})}\BibitemShut {NoStop}%
\bibitem [{\citenamefont {Andersson}\ \emph {et~al.}(2019)\citenamefont
  {Andersson}, \citenamefont {Suri}, \citenamefont {Guo}, \citenamefont
  {Aref},\ and\ \citenamefont {Delsing}}]{andersson2019non}%
  \BibitemOpen
  \bibfield  {author} {\bibinfo {author} {\bibfnamefont {G.}~\bibnamefont
  {Andersson}}, \bibinfo {author} {\bibfnamefont {B.}~\bibnamefont {Suri}},
  \bibinfo {author} {\bibfnamefont {L.}~\bibnamefont {Guo}}, \bibinfo {author}
  {\bibfnamefont {T.}~\bibnamefont {Aref}},\ and\ \bibinfo {author}
  {\bibfnamefont {P.}~\bibnamefont {Delsing}},\ }\href@noop {} {\bibfield
  {journal} {\bibinfo  {journal} {Nature Physics}\ }\textbf {\bibinfo {volume}
  {15}},\ \bibinfo {pages} {1123} (\bibinfo {year} {2019})}\BibitemShut
  {NoStop}%
\bibitem [{\citenamefont {Milonni}\ \emph {et~al.}(1983)\citenamefont
  {Milonni}, \citenamefont {Ackerhalt}, \citenamefont {Galbraith},\ and\
  \citenamefont {Shih}}]{milonni1983exponential}%
  \BibitemOpen
  \bibfield  {author} {\bibinfo {author} {\bibfnamefont {P.}~\bibnamefont
  {Milonni}}, \bibinfo {author} {\bibfnamefont {J.}~\bibnamefont {Ackerhalt}},
  \bibinfo {author} {\bibfnamefont {H.}~\bibnamefont {Galbraith}},\ and\
  \bibinfo {author} {\bibfnamefont {M.-L.}\ \bibnamefont {Shih}},\ }\href@noop
  {} {\bibfield  {journal} {\bibinfo  {journal} {Physical Review A}\ }\textbf
  {\bibinfo {volume} {28}},\ \bibinfo {pages} {32} (\bibinfo {year}
  {1983})}\BibitemShut {NoStop}%
\bibitem [{\citenamefont {Cohen-Tannoudji}\ \emph {et~al.}(1998)\citenamefont
  {Cohen-Tannoudji}, \citenamefont {Dupont-Roc},\ and\ \citenamefont
  {Grynberg}}]{cohen1998atom}%
  \BibitemOpen
  \bibfield  {author} {\bibinfo {author} {\bibfnamefont {C.}~\bibnamefont
  {Cohen-Tannoudji}}, \bibinfo {author} {\bibfnamefont {J.}~\bibnamefont
  {Dupont-Roc}},\ and\ \bibinfo {author} {\bibfnamefont {G.}~\bibnamefont
  {Grynberg}},\ }\href@noop {} {\emph {\bibinfo {title} {Atom-photon
  interactions: basic processes and applications}}}\ (\bibinfo  {publisher}
  {John Wiley \& Sons},\ \bibinfo {year} {1998})\BibitemShut {NoStop}%
\bibitem [{\citenamefont {Stey}\ and\ \citenamefont
  {Gibberd}(1972)}]{stey1972decay}%
  \BibitemOpen
  \bibfield  {author} {\bibinfo {author} {\bibfnamefont {G.}~\bibnamefont
  {Stey}}\ and\ \bibinfo {author} {\bibfnamefont {R.}~\bibnamefont {Gibberd}},\
  }\href@noop {} {\bibfield  {journal} {\bibinfo  {journal} {Physica}\ }\textbf
  {\bibinfo {volume} {60}},\ \bibinfo {pages} {1} (\bibinfo {year}
  {1972})}\BibitemShut {NoStop}%
\bibitem [{\citenamefont {Fano}(1935)}]{fano1935sullo}%
  \BibitemOpen
  \bibfield  {author} {\bibinfo {author} {\bibfnamefont {U.}~\bibnamefont
  {Fano}},\ }\href@noop {} {\bibfield  {journal} {\bibinfo  {journal} {Il Nuovo
  Cimento (1924-1942)}\ }\textbf {\bibinfo {volume} {12}},\ \bibinfo {pages}
  {154} (\bibinfo {year} {1935})}\BibitemShut {NoStop}%
\bibitem [{\citenamefont {Coish}\ \emph {et~al.}(2008)\citenamefont {Coish},
  \citenamefont {Fischer},\ and\ \citenamefont {Loss}}]{coish2008exponential}%
  \BibitemOpen
  \bibfield  {author} {\bibinfo {author} {\bibfnamefont {W.~A.}\ \bibnamefont
  {Coish}}, \bibinfo {author} {\bibfnamefont {J.}~\bibnamefont {Fischer}},\
  and\ \bibinfo {author} {\bibfnamefont {D.}~\bibnamefont {Loss}},\ }\href@noop
  {} {\bibfield  {journal} {\bibinfo  {journal} {Physical Review B}\ }\textbf
  {\bibinfo {volume} {77}},\ \bibinfo {pages} {125329} (\bibinfo {year}
  {2008})}\BibitemShut {NoStop}%
\bibitem [{\citenamefont {Micklitz}\ \emph {et~al.}(2022)\citenamefont
  {Micklitz}, \citenamefont {Morningstar}, \citenamefont {Altland},\ and\
  \citenamefont {Huse}}]{micklitz2022emergence}%
  \BibitemOpen
  \bibfield  {author} {\bibinfo {author} {\bibfnamefont {T.}~\bibnamefont
  {Micklitz}}, \bibinfo {author} {\bibfnamefont {A.}~\bibnamefont
  {Morningstar}}, \bibinfo {author} {\bibfnamefont {A.}~\bibnamefont
  {Altland}},\ and\ \bibinfo {author} {\bibfnamefont {D.~A.}\ \bibnamefont
  {Huse}},\ }\href@noop {} {\bibfield  {journal} {\bibinfo  {journal} {Physical
  review letters}\ }\textbf {\bibinfo {volume} {129}},\ \bibinfo {pages}
  {140402} (\bibinfo {year} {2022})}\BibitemShut {NoStop}%
\bibitem [{\citenamefont {Guo}\ \emph {et~al.}(2017)\citenamefont {Guo},
  \citenamefont {Grimsmo}, \citenamefont {Kockum}, \citenamefont {Pletyukhov},\
  and\ \citenamefont {Johansson}}]{guo2017giant}%
  \BibitemOpen
  \bibfield  {author} {\bibinfo {author} {\bibfnamefont {L.}~\bibnamefont
  {Guo}}, \bibinfo {author} {\bibfnamefont {A.}~\bibnamefont {Grimsmo}},
  \bibinfo {author} {\bibfnamefont {A.~F.}\ \bibnamefont {Kockum}}, \bibinfo
  {author} {\bibfnamefont {M.}~\bibnamefont {Pletyukhov}},\ and\ \bibinfo
  {author} {\bibfnamefont {G.}~\bibnamefont {Johansson}},\ }\href@noop {}
  {\bibfield  {journal} {\bibinfo  {journal} {Physical Review A}\ }\textbf
  {\bibinfo {volume} {95}},\ \bibinfo {pages} {053821} (\bibinfo {year}
  {2017})}\BibitemShut {NoStop}%
\bibitem [{\citenamefont {Yeh}\ \emph {et~al.}(1982)\citenamefont {Yeh},
  \citenamefont {Bowden},\ and\ \citenamefont {Eberly}}]{yeh1982interrupted}%
  \BibitemOpen
  \bibfield  {author} {\bibinfo {author} {\bibfnamefont {J.}~\bibnamefont
  {Yeh}}, \bibinfo {author} {\bibfnamefont {C.}~\bibnamefont {Bowden}},\ and\
  \bibinfo {author} {\bibfnamefont {J.}~\bibnamefont {Eberly}},\ }\href@noop {}
  {\bibfield  {journal} {\bibinfo  {journal} {The Journal of Chemical Physics}\
  }\textbf {\bibinfo {volume} {76}},\ \bibinfo {pages} {5936} (\bibinfo {year}
  {1982})}\BibitemShut {NoStop}%
\bibitem [{\citenamefont {Eberly}\ \emph {et~al.}(1980)\citenamefont {Eberly},
  \citenamefont {Narozhny},\ and\ \citenamefont
  {Sanchez-Mondragon}}]{eberly1980periodic}%
  \BibitemOpen
  \bibfield  {author} {\bibinfo {author} {\bibfnamefont {J.~H.}\ \bibnamefont
  {Eberly}}, \bibinfo {author} {\bibfnamefont {N.}~\bibnamefont {Narozhny}},\
  and\ \bibinfo {author} {\bibfnamefont {J.}~\bibnamefont
  {Sanchez-Mondragon}},\ }\href@noop {} {\bibfield  {journal} {\bibinfo
  {journal} {Physical Review Letters}\ }\textbf {\bibinfo {volume} {44}},\
  \bibinfo {pages} {1323} (\bibinfo {year} {1980})}\BibitemShut {NoStop}%
\bibitem [{\citenamefont {Gaveau}\ and\ \citenamefont
  {Schulman}(1995)}]{gaveau1995limited}%
  \BibitemOpen
  \bibfield  {author} {\bibinfo {author} {\bibfnamefont {B.}~\bibnamefont
  {Gaveau}}\ and\ \bibinfo {author} {\bibfnamefont {L.}~\bibnamefont
  {Schulman}},\ }\href@noop {} {\bibfield  {journal} {\bibinfo  {journal}
  {Journal of Physics A: Mathematical and General}\ }\textbf {\bibinfo {volume}
  {28}},\ \bibinfo {pages} {7359} (\bibinfo {year} {1995})}\BibitemShut
  {NoStop}%
\bibitem [{\citenamefont {Fano}(1961)}]{fano1961effects}%
  \BibitemOpen
  \bibfield  {author} {\bibinfo {author} {\bibfnamefont {U.}~\bibnamefont
  {Fano}},\ }\href@noop {} {\bibfield  {journal} {\bibinfo  {journal} {Physical
  review}\ }\textbf {\bibinfo {volume} {124}},\ \bibinfo {pages} {1866}
  (\bibinfo {year} {1961})}\BibitemShut {NoStop}%
\bibitem [{Note1()}]{Note1}%
  \BibitemOpen
  \bibinfo {note} {This discussion assumes that the revival peaks are
  completely suppressed when the next revival time sets in. In the opposite
  case on should expect further interference effects between contributions from
  different revival-time intervals.}\BibitemShut {Stop}%
\bibitem [{\citenamefont {Dorner}\ and\ \citenamefont
  {Zoller}(2002)}]{dorner2002laser}%
  \BibitemOpen
  \bibfield  {author} {\bibinfo {author} {\bibfnamefont {U.}~\bibnamefont
  {Dorner}}\ and\ \bibinfo {author} {\bibfnamefont {P.}~\bibnamefont
  {Zoller}},\ }\href@noop {} {\bibfield  {journal} {\bibinfo  {journal}
  {Physical Review A}\ }\textbf {\bibinfo {volume} {66}},\ \bibinfo {pages}
  {023816} (\bibinfo {year} {2002})}\BibitemShut {NoStop}%
\bibitem [{Note2()}]{Note2}%
  \BibitemOpen
  \bibinfo {note} {This becomes clear from the simpler polynomial structure
  accompanying the exponentials in the sum.}\BibitemShut {Stop}%
\bibitem [{\citenamefont {Gradshteyn}\ and\ \citenamefont
  {Ryzhik}(1965)}]{Gradshteyn:1943cpj}%
  \BibitemOpen
  \bibfield  {author} {\bibinfo {author} {\bibfnamefont {I.~S.}\ \bibnamefont
  {Gradshteyn}}\ and\ \bibinfo {author} {\bibfnamefont {I.~M.}\ \bibnamefont
  {Ryzhik}},\ }\href@noop {} {\emph {\bibinfo {title} {{Table of Integrals,
  Series, and Products}}}}\ (\bibinfo  {publisher} {Academic Press},\ \bibinfo
  {year} {1965})\BibitemShut {NoStop}%
\end{thebibliography}%

\end{document}